\documentclass[prd,preprint,aps]{revtex4}

\usepackage{graphicx}
\usepackage{latexsym}
\usepackage{amsmath}
\usepackage{amsfonts}
\usepackage{amssymb}

\newcommand{\be}{\begin{equation}}
\newcommand{\ee}{\end{equation}}
\newcommand{\ben}{\begin{eqnarray}}
\newcommand{\een}{\end{eqnarray}}
\newcommand{\sech}{\rm sech}

\begin{document}

\title{Accelerating universes driven by bulk particles}

\author{F.A. Brito$^{a}$, F.F. Cruz$^{b}$
and J.F.N. Oliveira$^{b}$}

\affiliation{$^a$Departamento de F\'\i sica, Universidade Federal
de Campina Grande, 58109-970 Campina Grande, Para\'\i ba, Brazil
\\
$^b$Departamento de Matem\'atica, Universidade Regional do Cariri,
63040-000 Juazeiro do Norte, Cear\'a, Brazil}

\date{\today}

\begin{abstract}
We consider our universe as a 3d domain wall embedded in a 5d
dimensional Minkowski space-time. We address the problem of
inflation and late time acceleration driven by bulk particles
colliding with the 3d domain wall. The expansion of our universe
is mainly related to these bulk particles. Since our universe
tends to be permeated by a large number of isolated structures, as
temperature diminishes with the expansion, we model our universe
with a 3d domain wall with increasing internal structures. These
structures could be unstable 2d domain walls evolving to
fermi-balls which are candidates to cold dark matter. The momentum
transfer of bulk particles colliding with the 3d domain wall is
related to the reflection coefficient. We show a nontrivial
dependence of the reflection coefficient with the number of
internal dark matter structures inside the 3d domain wall. As the
population of such structures increases the velocity of the domain
wall expansion also increases. The expansion is exponential at
early times and polynomial at late times. We connect this picture
with string/M-theory by considering BPS 3d domain walls with
structures which can appear through the bosonic sector of a
five-dimensional supergravity theory.
\end{abstract}


\maketitle

\section{Introduction}

Following the recent ideas concerning extra-dimensions
\cite{nima1,nima2,rs1,rs2,kr} we consider our universe as a 3d
domain wall embedded in a 5d dimensional space-time. We explore some
current issues of the modern cosmology such as inflation and late
time accelerating universes. Accelerating expansion can be modelled
by considering a dynamical cosmological constant $\Lambda$
\cite{peebles,steinhardt} in the form of a scalar field with a
scalar potential and a slowly varying energy density. This idea is
currently called {\it quintessence}. In this paper we show how this
mechanism can appear via bulk particles collisions with domain
walls. These are 3d domain walls engendering an amount of internal
structures increasing with time. Such structures can be unstable 2d
domain walls evolving to fermi-balls which are candidates to cold
dark matter \cite{macpherson,bazmorris}. This is an interesting
alternative to understand acceleration of our universe as a
consequence of bulk effects. This can be relevant to reproduce
accelerating universes in string/M-theory cosmology
\cite{vijay,kklt,kklmmt}. Accelerating universes require violation
of the strong energy condition that is not allowed by potentials
obtained through compactifications from ten/eleven-dimensional down
to four-dimensional supergravity theories \cite{nogoth1}
$^{[1]}$\footnotetext[1]{Indeed, time dependent compactifications
\cite{townsend0,ohta} provide cosmologies with transient phase of
acceleration, without future event horizon \cite{townsend}.}.
However, these potentials can produce BPS solitonic solutions such
as a BPS 3d domain wall (3-brane) inside a bulk \cite{nunez} whose
dynamics between the 3-brane and particles inside the bulk can play
the crucial role of giving rise to accelerating universes on a
3-brane. In other words, a {\it brane observer} can ``see'' a
potential on a BPS 3-brane induced by bulk particles collisions.
This potential unlike the potentials coming from compactifications
can violate the strong energy condition and then can be suitable to
realize accelerating universes. In this scenario the dark energy
effect needed to drive both early and late time acceleration can be
recognized as the effect of ``brane dark matter'' plus ``bulk
particle collisions.'' Such particles travelling the bulk can be
associated with the 3-brane fluctuations effect itself or some
exotic particles. We show that this scenario can also accommodate
inflation in early time cosmology. Our scenario can be considered as
a new perspective to accelerating universes in braneworlds cosmology
scenarios \cite{dgp,dg,defayet,ac-davis}.

In this paper we follow a program we have considered in several
contexts
\cite{mackenzie,morris97,bb97,etbb97,bbb,bazbrito,bb2000,bb2001}.
This program concerns the issue of localization of defects inside
other defects of higher dimensions in field theory. This was
inspired by the idea of superconducting cosmic strings
\cite{witten85}. Essentially one considers the fact that inside,
i.e., near the core of a defect, a scalar or a vector field can
develop a nonzero vacuum expectation value (v.e.v.). To provide
structures formation such as domain walls inside another domain wall
one has to consider a theory with a $Z_2\times Z_2$ symmetry. Thus,
in our analysis, the appearance of a scalar field developing a
v.e.v. inside a 3d domain wall, breaks a particular $Z_2$ symmetry
favoring the formation of another domain wall with smaller
dimension. These configurations of defects inside defects in field
theory have some analog in braneworlds with warped geometry where
one considers 3-branes inside higher dimensional spaces such as 5d
anti-de-Sitter space-time ($AdS_5$) \cite{rs1,rs2,kr}.

We shall consider here a 3d domain wall (a thick 3-brane) with
internal structure such as other domain walls. We mainly address
the problem of particle collision with a 3d domain wall embedded
in 5d Minkowski space-time \cite{dgp,dg}. We calculate the
reflection probability of these particles by considering they
collide elastically with the domain wall. As we shall see, the
total mass of the domain wall does not change as the number of
internal structure increases. Thus we disregard absorbtion of bulk
particles and also consider these particles are hitting the 3d
domain wall orthogonally, i.e., there is no tangential momentum
transfer to any individual structure inside the domain wall
$^{[2]}$\footnotetext[2]{Indeed, inflow/outflow of matter or
energy between the 3-brane and the bulk is not enough to
accelerate universes, because the flow is cancelled by ``mirage
effects'' \cite{tetradis1,tetradis2}. We thank N. Tetradis for
bringing this into our attention.}. In our scenario the domain
wall expands as it moves along the 5d space-time, and this motion
is essentially due to the total momentum transferred by bulk
particles. We consider two types of 3d domain wall solutions: (i)
domain walls without internal structure and (ii) domain walls with
internal structure. They appear as BPS solitonic solutions of the
bosonic sector of a 5d supergravity theory.

We argue that 3d domain walls with internal structure colliding
with bulk scalar particles can drive inflation in early time
cosmology and accelerating expansion in late time universe.
(Investigations concerning accelerating universes in scenarios
with S-branes inside D-branes have been considered in
\cite{edels}). The expansion is controlled by the amount of
internal structures that increases with the time according to a
power-law. At early times the reflection probability increases
exponentially and then provides an inflationary period of our
universe. At very late times the reflection probability saturates
to a constant leading to a universe expanding according to a
quadratic-law. Accelerating expansion in our universe is in
agreement with recent cosmological observational data
\cite{supernovae,Ries}.

The paper is organized as in the following. In Sec.~\ref{model} we
find 2d domain walls inside a 3d domain wall as solitonic
solutions of the bosonic sector of a 5d supergravity theory. In
Sec.~\ref{schr} we present the quantum mechanics problem of bulk
particles interacting with Schroedinger potentials associated with
the 3d domain walls. In Sec.~\ref{refp} we calculate the
reflection probability and identify the number of 2d domain walls
as a parameter of control. The 3d domain wall expansion is
explored in detail in Sec.~\ref{DWE}. Final discussions are
considered in Sec.~\ref{conclu}.

\section{The model: walls inside wall}
\label{model}

We consider the bosonic scalar sector of a five-dimensional
supergravity theory obtained via compactification of a higher
dimensional supergravity whose Lagrangian is
\cite{cv,dwolfe,bcy,cvlambert,bbn} \ben \label{sugra} e^{-1}{\cal
L}_{sugra}=-\frac{1}{4}\,{M_*}^3R_{(5)}+\,G_{AB}\,\partial_\mu\phi^A\,\partial^\mu\phi^B
-\frac{1}{4}\,G^{AB}\,\frac{\partial
W(\phi)}{\partial\phi^A}\frac{\partial
W(\phi)}{\partial\phi^B}+\frac{1}{3}\frac{1}{{M_*}^3}\,W(\phi)^2,
\een where $G_{AB}$ is the metric on the scalar target space and
$e=|\det g_{\mu\nu}|^{1/2}$. $R_{(5)}$ is the 5d Ricci scalar and
$1/M_*$ is the five-dimensional Planck length. Below we consider
the limit $M_*\!\gg\!1$ (with $M_*\ll M_{Pl}$) where
five-dimensional gravity is not coupled to the scalar field. In
this limit the 3-brane is a classical 3d domain wall solution
embedded in a 5d Minkowski space. The 4d gravity is induced on the
brane via quantum loops of matter fields confined on the brane
\cite{dgp,dg}. The Standard Model particles are localized on the
brane via mechanisms described in \cite{nima1,nima2} --- see also
\cite{susy1SM} for realizations in string models. We also restrict
the scalar manifold to two fields only, i.e.,
$\phi_A=(\phi/\sqrt{2},\chi/\sqrt{2})$.

In order to have formation of domain walls inside domain walls, we
consider a supersymmetric model with two coupled real scalar fields
engendering a $Z_2 \times Z_2$ symmetry (for now we consider this as
an exact symmetry) described by the scalar sector of the Lagrangian
(\ref{sugra}), i.e., \be\label{01}
\mathcal{L}=\frac{1}{2}\partial_{\mu}\phi\partial^{\mu}\phi+\frac{1}{2}
\partial_{\mu}\chi\partial^{\mu}\chi-V(\phi,\chi), \ee where the scalar
potential is given in terms of the superpotential as \be\label{02}
V(\phi,\chi)= \frac{1}{2}\left(\frac{\partial
W}{\partial\phi}\right)^2+\frac{1}{2}\left(\frac{\partial
W}{\partial\chi}\right)^2.\ee The most general form of a
superpotential engendering a $Z_2 \times Z_2$ symmetry \cite{bbb} is
\be\label{03} W=
\lambda\left(\frac{\phi^3}{3}-a^2\phi\right)+\mu\phi\chi^2.\ee The
physical meaning of the parameters $\lambda$ and $\mu$ will become
clear later as we recognize $N=\sqrt{\mu/\lambda}$ as the number of
internal structures inside the 3d domain wall. Using the
superpotential above gives
\be\label{05}V(\phi,\chi)=\frac{1}{2}\lambda^2(\phi^2-a^2)^2+(2\mu^2+
\lambda\mu)\phi^2\chi^2-\lambda\mu
a^2\chi^2+\frac{1}{2}\mu^2\chi^4.\ee The equations of motion are
\ben\label{f3}
\partial_{\mu}\partial^{\mu}\phi+\frac{\partial
V}{\partial\phi}=0
\\ \label{f4}
\partial_{\mu}\partial^{\mu}\chi+\frac{\partial
V}{\partial\chi}=0.\een

\begin{figure}[ht]
\centerline{\includegraphics[{angle=90,height=7.0cm,angle=180,width=8.0cm}]
{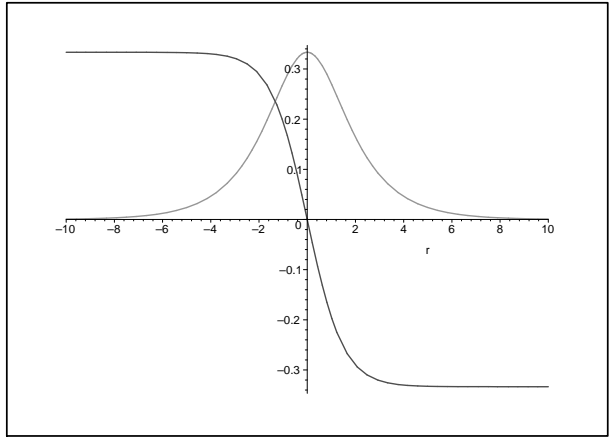}} \caption{The field $\chi$ is localized inside
($\phi\approx0$) the $\phi$-domain wall (anti-kink).}\label{fig3}
\end{figure}

For simplicity we look for one-dimensional static soliton
solutions. By considering $\phi\equiv\phi(r)$ and
$\chi\equiv\chi(r)$ the equations of motion (\ref{f3}) e
(\ref{f4}) are given by
\ben\label{18}\frac{d^2\phi}{dr^2}&=&2\lambda^2(\phi^2-a^2)\phi+
(4\mu^2+2\lambda\mu)\phi\chi^2 \\
\label{19}\frac{d^2\chi}{dr^2}&=&2\mu^2(\chi^2-\frac{\lambda}{\mu}a^2)\chi+
(4\mu^2+2\lambda\mu)\phi^2\chi\een Since this is a supersymmetric
system, we can investigate this problem by using the first-order
equation formalism \ben\label{20}\frac{d\phi}{dr} &=&
W_{\phi}=\lambda(\phi^2-a^2)+\mu\phi\chi^2
\\ \label{21}\frac{d\chi}{dr} &=& W_{\chi}=2\mu\phi\chi \een
where the subscripts $\phi,\,\chi\,...$ mean derivatives with
respect to these fields. The solutions of the first-order
equations above are the well-known BPS solutions
\cite{bazmauro,brs96,bnrt,shv,bazbrito}

\noindent i) Type I:
\ben\label{22}\phi &=&-a\tanh(\lambda ar) \\
\nonumber \chi &=& 0\een \noindent ii) Type II: \ben\label{23}\phi
&=& -a\tanh(2\mu ar)
\\ \nonumber \chi &=&\pm a\sqrt{\frac{\lambda}{\mu}-2}\,{\sech}(2\mu
ar),\een where $r$ is a fifth coordinate transverse to the 3d
domain wall. The first pair of solutions represents a domain wall
with no structure, whereas the second pair represents a domain
wall with internal structure. The latter case has been consider in
\cite{bgomes} in the context of Bloch brane scenario. In both
configurations the system has the same Bogomol'nyi energy
$E_B=(4/3)\lambda a^3$ (i.e., the domain wall mass $M_{wall}\equiv
E_B$), despite of the fact that in the second case one has
internal structures while in the first case one has not. In the
solution of the type II, as the field $\phi$ approaches to zero at
the core of the $\phi$-domain wall, $\phi(z\approx0)\approx0$, the
field $\chi$ develops its maximal value
--- see Fig.~\ref{fig3}. In this regime the potential
$V(\phi,\chi)$ given in (\ref{05}) approaches effectively to the
potential \ben \label{pot_eff}
V_{eff}(\chi)=\frac{1}{2}\mu^2\left(\chi^2-\frac{\lambda}{\mu}
a^2\right)^2. \een So we can conclude the dynamics inside the
$\phi$-domain wall ($\phi\approx0$) is governed by the effective
Lagrangian \be\label{01.2}
\mathcal{L}_{eff}=\frac{1}{2}\partial_{\mu}\chi\partial^{\mu}\chi
-V_{eff}(\chi).\ee Given the potential (\ref{pot_eff}), this
Lagrangian can describe other domain walls. We can represent a
particular domain wall by the kink/anti-kink solutions \ben
\label{parede_chi} \chi=\pm
a\sqrt{\frac{\lambda}{\mu}}\tanh{\left(\mu\sqrt{\frac{\lambda}{\mu}}x_i\right)}
\een where $x_i$ stands for some coordinate extended along the
$\phi$-domain wall. The 2d domain walls described by
(\ref{parede_chi}) are referred to as ``$\chi$-domain walls''. To
avoid the dangerous $\chi$-wall-dominated universe on the 3d
domain wall we assume an approximated $Z_2$-symmetry in the
effective Lagrangian (\ref{01.2}). Upon this circumstance the 2d
$\chi$-walls tend to collapse and then stabilize through trapping
of fermion zero modes to become stable fermi-balls
\cite{morris97,bb97}, which are candidates to cold dark matter
\cite{macpherson,bazmorris}.

As we shall see, a scalar meson interacting with domain walls
described by solutions of type II (\ref{23}) can feel the degree
of structure inside these domain walls, although they have the
same energy of the domain walls without structure given by
solutions of type I (\ref{22}). This suggests the whole energy of
the configuration type I appears distributed in building blocks in
the configuration type II. Such building blocks are internal
structures made of 2d $\chi$-domain walls given by
(\ref{parede_chi}).

\section{Schroedinger-like equation for the fluctuations}
\label{schr}

Let us consider linear fluctuations around the solutions ${\phi}$
and ${\chi}$. Consider $\chi=\bar{\chi}+\zeta$ and
$\phi=\bar{\Phi}$ where $\zeta$ is the fluctuation of the field
$\chi$, $\bar{\Phi}$ and $\bar{\chi}$ are solutions of equations
of motion (\ref{f3}) and (\ref{f4}), respectively. Expanding the
equations of motion around the fluctuations we find
\be\label{15}\partial_{\mu}\partial^{\mu}\zeta+
\bar{V}_{\chi\chi}\zeta=0.\ee We consider the following Ansatz for
the perturbation around a 3d domain wall extended along the
coordinates $(x,y,z)$ \be\label{16}\zeta=\zeta(r)e^{-i(\omega
t-k_xx-k_yy-k_zz)}.\ee Now substituting this equation into
(\ref{15}) we have
\be\label{17}-\partial_r^2\zeta(r)+[-\omega^2+k^2_x+k^2_y+k^2_z]\zeta(r)
+\bar{V}_{\chi\chi}\zeta(r)=0,\ee which can be identified with a
Schroedinger-like equation. We are interested in solving the
quantum mechanics problem to obtain the reflection coefficient of
particles as they reach the $\phi$-~3d~domain wall of type II. We
are only considering elastic collisions and that the particles are
hitting the 3d domain wall orthogonally.

Let us consider the Schroedinger potential for the case I, i.e.,
the 3d domain wall without internal structure. Using equation
(\ref{05}) and the solutions given in (\ref{22}) we have
\be\label{25}
\bar{V}_{\chi\chi}(r)=4\mu^2a^2-4\mu^2a^2(1+\frac{\lambda}{2\mu})\,{\sech}^2(\lambda
ar).\ee Note that as $\chi=0$ one finds $\phi=\pm a$ (``far'' from
the wall). In this limit,
$\bar{V}_{\chi\chi}(\pm\infty)=4\mu^2a^2$ is the mass of
$\chi$-particles outside the wall, i.e., $m_{\chi}^2=4\mu^2a^2$,
where we assume $\chi$ as being the field that describes the
particles interacting with the field $\phi$. Thus we have the
Schroedinger potential
\be\label{26}\bar{V}_{\chi\chi}(r)=m_{\chi}^2-m_{\chi}^2(1+\frac{\lambda}{2\mu})
\,{\sech}^2(\lambda ar).\ee It is useful to rewrite the righthand
side of this equation as $m^2_{\chi}+U(r)$, such that we can write
(\ref{17}) as
\be\label{27}-\partial_r^2\zeta(r)+[-\omega^2+k^2_x+k^2_y+k^2_z+m^2_{\chi}]
\zeta(r)+U(r)\zeta(r)=0.\ee Setting
$-k^2_r=-\omega^2+k_x^2+k_y^2+k_z^2+m_\chi^2$ we can write
(\ref{27}) simply as
\be\label{28}\zeta^{''}(r)+k^2_r\zeta(r)-U_I(r)\zeta(r)=0,\ee
where $k_r$ is the $r$-component of a bulk particle momentum and
\be\label{29}U_I(r)=-m^2_{\chi}(1+\frac{\lambda}{2\mu})\,{\sech}^2(\lambda
ar).\ee

Finally let us consider the Schroedinger potential for the case
II, i.e., the 3d domain walls with internal structure. Using
(\ref{05}) and the solutions given in (\ref{23}) we have
\be\label{31}\bar{V}_{\chi\chi}(r)=4\mu^2a^2-4\mu^2a^2(4-\frac{\lambda}{\mu})
\,{\sech}^2(2\mu ar),\ee with $m_{\chi}^2=4\mu^2a^2$ and
\be\label{32}\bar{V}_{\chi\chi}(r)=m^2_{\chi}-m^2_{\chi}(4-\frac{\lambda}{\mu})
\,{\sech}^2(2\mu ar).\ee Just as in the previous case we have
\be\label{33}\zeta^{''}(r)+k^2_r\zeta(r)-U_{II}(r)\zeta(r)=0,\ee
where the Schroedinger potential now reads
\be\label{34}U_{II}(r)=-m^2_{\chi}(4-\frac{\lambda}{\mu})\,{\sech}^2(2\mu
ar).\ee We note that in both cases (\ref{29}) and (\ref{34}) we
have the same $\chi$-particles with squared masses $m_{\chi}^2$.
However, as $U_I$ has width $1/\lambda a$ and depth
$-m_{\chi}^2(1+\lambda/2\mu)$ whereas $U_{II}$ has width $1/2\mu
a$ and depth $-m_{\chi}^2(4-\lambda/\mu)$ the $\chi$-mesons
clearly should feel different effects.

\section{Reflection Probability}
\label{refp}

The reflection coefficient for potentials like
$U(r)=-U_0\,{\sech}^2{\alpha r}$ is given by \cite{vilenkin} \ben
\label{R}R=\frac{\cos^2{\beta}}{\sinh^2{\gamma}+\cos^2{\beta}}
\een with $\gamma=\pi k_r/\alpha$ and
$\beta=(\pi/2)\sqrt{1+4U_0/\alpha^2}$. Let us now focus on the
physics of the case II. Thus for the potential $U_{II}$ we have
\ben \label{RII}
R=\frac{\cos^2{\left(\frac{\pi}{2}\sqrt{17-\frac{4}{N^2}}\,\right)}}{\sinh^2{\left(\frac{\pi
k_r\delta}{N^2}\right)}+\cos^2{\left(\frac{\pi}{2}\sqrt{17-\frac{4}{N^2}}\,\right)}}
\een where we identify $N=\sqrt{\mu/\lambda}$ with the number of
$\chi$-~2d~domain walls inside the $\phi$-~3d~domain walls that
have width $\simeq\delta=1/\lambda a$; and $k_r=2\pi/\lambda'$ is
the $r$-component of the wave propagation vector associated with a
bulk $\chi$-particle. The estimative about the number $N$ is made
by considering the fact that both solutions of type I and II have
the same energy $E_B^{\phi(I,II)}=(4/3)\lambda a^3$, as we
mentioned earlier. Note also that the solution of type II can be
deformed to the solution of type I by setting $\lambda/\mu=2$. By
distributing the whole energy of the $\phi$-3d domain wall of the
case II in terms of energy of the internal structures, i.e.,
$\chi$-domain walls, we have $E_B^{\phi(II)}=NE_B^\chi$, where
$E_B^\chi=(4/3)\lambda a^3\sqrt{\lambda/\mu}$ is the energy of a
$\chi$-domain wall. This clearly reproduce the value of $N$
aforementioned.

\begin{figure}[ht]
\centerline{\includegraphics[{angle=90,height=7.0cm,angle=180,width=8.0cm}]
{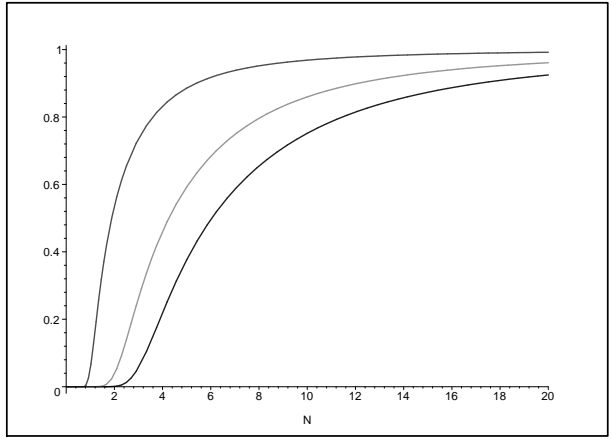}} \caption{Reflection coefficient $R$ as
function of the number of $\chi$-domain walls for a bulk
$\chi$-particle colliding with a 3d domain wall with $k_r=1,\,5$
and 10 ($\delta=1$).}\label{fig1}
\end{figure}

On the one hand, if the wavelength $\lambda'$ of the
$\chi$-particles is much smaller than the width $\delta$ of the
$\phi$-domain wall, thus $k_r\gg(\delta^{-1}\!=\!\lambda a)$ and
$R$ is exponentially small. On the other hand, if $k_r\to0$,
$R\to1$ and almost all $\chi$-particles are reflected. In
Fig.~\ref{fig1} we see $R$ as a function of the number $N$ of
$\chi$-domain walls. Note also that for $\lambda/\mu=2$ ({\it
domain wall without internal structure}) we find $1/N^2=2$ and
then $R=0$ which implies that {\it there is no reflection}.

\begin{figure}[ht]
\centerline{\includegraphics[{angle=90,height=7.0cm,angle=180,width=8.0cm}]
{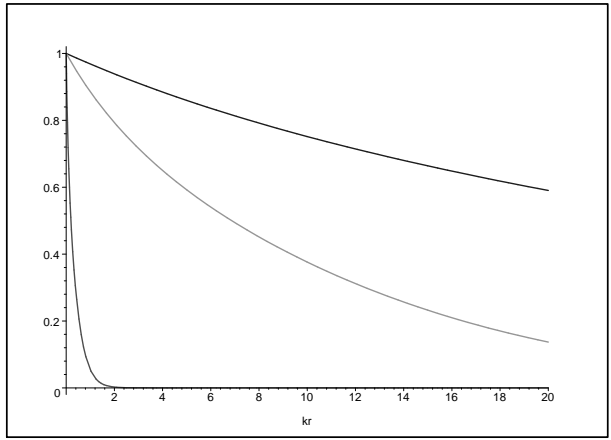}} \caption{Reflection coefficient $R$ as a
function of the momentum $k_r$ of a bulk $\chi$-particle colliding
with a 3d domain wall with $N=1,\,5$ and 10 $\chi$-domain walls
($\delta=1$).}\label{fig2}
\end{figure}

The behavior of the reflection coefficient can also be seen as a
function of the momentum $k_r$ of a $\chi$-particle colliding with
a $\phi$-3d domain wall fulfilled of $\chi$-domain walls --- see
Fig.~\ref{fig2}.

\section{Domain Wall Expansion}
\label{DWE}

In this section we investigate how the 3d domain wall expands
along its directions as bulk $\chi$-particles collide with it. We
use the reflection probability to calculate the rate of momentum
transfer of such particles to the domain wall. We can estimate the
domain wall acceleration along the transverse coordinate $r$ due
to elastic collisions as a function of the reflection coefficient
$R$ via transversal force per unit area \ben \label{acceler}
F_r=M_{wall}\,\ddot{r}(t)\simeq\kappa R,\een where $r(t)$ is the
position (or ``radius'') of the domain wall with respect to the
origin $O$ into the bulk; $\kappa$ has dependence on the density
of colliding bulk particles and their incoming momenta $k_r$, that
we assume to be time independent.

To relate the size of the coordinates along the domain wall with
its position $r(t)$ into the bulk $^{[3]}$\footnotetext[3]{A
similar set up is found in the {\it brane inflation} scenario
\cite{dtye}.} we have to make some assumptions. To guide ourselves
let us first consider a simple example: a spherical 2d domain wall
separating the bulk into two asymmetric regions. The line element
on the 2d domain wall depends on $r(t)$ as
$ds_2^2=r(t)^2(d\theta^2+\sin^2{\theta}d\varphi^2)$; where
$(\theta,\varphi)$ are angular coordinates on the surface of the
2d domain wall. In terms of co-moving coordinates we define
$\vec{r}(t)=a(t)\vec{r}$ and the line element becomes
$ds_2^2=a(t)^2r^2(d\theta^2+\sin^2{\theta}d\varphi^2)=a(t)^2(dx^2+dy^2)$,
where $(x,y)$ are co-moving coordinates. The generalization to a
3d domain wall is straightforward:
$ds_3^2=a(t)^2(dx^2+dy^2+dz^2)$. Another example is considering a
domain wall as the boundary of a cone whose size of the domain
wall varies with the size of the cone. In our analyze below we
consider only the former example.

Until now we have considered a flat 3d domain wall. Indeed, the
analysis above is also valid for spherical walls as long as we are
considering a regime where the wall radius is very large compared
with the wall thickness --- {\it thin wall limit}. Curved domain
walls can be realized in field theory by breaking the
$Z_2$-symmetry slightly via inclusion of a very small term, e.g.,
$\epsilon\phi^3$, into the original Lagrangian
--- see \cite{kolb}. On this account let us
consider our 4d universe as a 3d domain wall evolving in time and
identify $\vec{r}(t)=a(t)\vec{x}$ with the radius of an expanding
3-sphere whose metric is \ben \label{ds}ds_4^2=dt^2
-a(t)^2(dx^2+dy^2+dz^2),\een where $a(t)$ is the scale factor of
our universe and $\vec{x}=(x,y,z)$ are co-moving coordinates.

Now let us consider two important limits. By considering the
number of internal structure increases with the time according to
the law $N\sim(\beta t)^\alpha$ ($\alpha\geq1$) and substituting
into (\ref{acceler}) we get to the power-law expansion of an
accelerating 3d domain wall \ben
\label{expansion}a(t)\sim\frac{1}{2}\kappa t^2, \een which is the
leading term for large times, i.e., $\beta t\gg k_r \delta$. This
means an accelerating universe at late time cosmology.  This fact
agrees with current observations \cite{supernovae,Ries}.

Assuming the expansion of our universe, for a brane observer, is
driven by an induced scalar field $\varphi$, confined on the 3d
domain wall $^{[4]}$\footnotetext[4]{The 4d gravity is also
induced on the brane via quantum loops of the matter fields on the
brane \cite{dgp,dg}.}, depending only on the time coordinate, the
$\varphi$ e.o.m and the Friedmann equation are given by \ben
\label{scalar}
\ddot\varphi+3H\dot\varphi=-\frac{\partial V}{\partial\varphi},\\
\label{Fdman}3H^2\equiv 3\left(\frac{\dot{a}}{a}\right)^2=
\frac{1}{2}\dot\varphi^2+V(\varphi). \een  For exponential
potentials as \ben \label{expV} V(\varphi)\sim V_0
e^{-2\frac{\varphi}{\varphi_0}}, \een which have been much
explored in string/M-theory context \cite{odintsov,ac-davis}, one
can find solutions satisfying (\ref{scalar}) and (\ref{Fdman}):
\ben \label{solCosmos}
a(t)=a_0\left({\beta}{t}\right)^{\frac{1}{2}\varphi_0^2}, \qquad
\varphi=\varphi_0\ln{\left({\beta}{t}\right)},\een where \ben
\beta^{-1}=\varphi_0\sqrt{\frac{1}{V_0}\left(\frac{3}{4}\varphi_0^2
-\frac{1}{2}\right)}.\een See, e.g., \cite{s.sen} for other
examples of scalar potentials. Comparing equation
(\ref{solCosmos}) with the scale factor $a(t)\sim t^2$ given in
(\ref{expansion}) we find $\varphi_0=2$ and
$\beta^{-1}=\sqrt{10/V_0}$. A necessary requirement for the
observed acceleration in the present universe is a power-law
expansion $a(t)\sim t^{\alpha}$ with $\alpha>1$.

In the early time cosmology, inflation is a crucial phenomenon for
solving several problems such as flatness of universe, monopole
problem, etc. At this stage the universe expands according to an
exponential-law $a(t)\sim e^{\alpha t}$ ($\alpha>0$). In our
scenario for very early times $\beta t\ll k_r \delta$ the
acceleration is such that the expansion behaves as
\ben\label{acelera} {a}(t)\simeq\frac{1}{2}\kappa\,\beta^2 t^2
e^{-\frac{\alpha}{\beta^2 t^2}},\een which can be well
approximated by an exponential-law $e^{\sqrt{\Lambda_0} t}$, for
$\alpha=2\pi k_r\delta$ small enough (i.e., $k_r\delta$ small,
that means bulk particles with low momenta $k_r$). This
inflationary phase starts for $N^2>1/2$, i.e., as the number of
internal structure becomes to play a fundamental role in the 3d
domain wall expansion. This corresponds to the time $t> (1/\beta)
(1/\sqrt{2})^{1/\alpha}$. Thus our scenario also reproduces the
inflationary phase in the early time universe. The evolution of a
dynamical cosmological constant \cite{peebles,steinhardt} on the
3d domain wall can be given by the Friedmann equation \ben
\label{Lambda}
\left(\frac{\dot{a}}{a}\right)^2\simeq\frac{\Lambda(t)}{3},\een
whose time dependence is
$\Lambda(t)\simeq12(\beta^2t^2+\alpha)^2/t^6\beta^4$. At early
times ($\beta t\ll1$) the cosmological constant becomes large,
$\Lambda_0\simeq 12\alpha^2/\beta^4t^{6}$, and dominant in
(\ref{Lambda}). As time goes larger the cosmological constant
approaches to zero according to the power-law
$\Lambda\simeq12/t^2$. This goes like the exponent of the
exponential factor in (\ref{acelera}). For time large enough we
should find a transition where the exponential factor is
suppressed and the quadratic pre-factor in (\ref{acelera})
reproduces the late time cosmology formula (\ref{expansion}). If
we consider this transition occurring around a time equals to the
observed age of our universe, i.e., $t\sim15$ billion of years we
get to the cosmological constant
\ben\label{Lambda0}\Lambda\sim10^{-34}s^{-2}\sim10^{-119}M_{Pl}^2
.\een Or in terms of 4d Planck length $L_{Pl}=1/M_{Pl}\equiv(8\pi
G_N)^{1/2}$ we define the dimensionless quantity $\Lambda L_{Pl}^2
\sim10^{-119}$. According to the analysis above we get to the
power-law $\Lambda\sim1/t^2$. Thus we can infer that there is a
tendency of having a {\it null} cosmological constant in our
universe for a very long time. Following the philosophy of
dynamical cosmological constant \cite{peebles,steinhardt}, or
simply quintessence, we could argue that the ``observed''
cosmological constant assumes the value given in (\ref{Lambda0})
because of the current age of our universe. Although this idea is
not new, here it seems to be applicable in a very natural way.
This is because the dynamical cosmological constant arises due to
the physical effect of bulk particles colliding with a 3d domain
wall (our universe), instead arising from a fine-tuned scalar
potential --- see \cite{weinberg,padna} for a review on the
cosmological constant problems.

\section{Discussions}
\label{conclu}

In early times before inflation we have radiation dominance on the
3d domain wall. The expansion follows according to the power-law
$a(t)\sim t^{1/2}$ and temperature decreases as $T\sim a(t)^{-1}$.
As temperature decreases, symmetry breaking inside the 3d domain
wall provides formation of internal structure favoring momentum
transfer of bulk particles to the domain wall. Particles with
larger momenta can be mainly reflected only later by a large
amount of internal structures on the 3d domain wall while
particles with lower momenta can be reflected by just a few number
of internal structure and then they can transfer momentum to the
domain wall earlier. Thus particles carrying low momenta
``ignite'' the inflationary expansion in its very beginning. Once
this expansion is started the cooling process on the 3d domain
wall favors even more the formation of internal structure with the
time according to a power-law. At this time, particles with larger
momenta also turn to participate of the process. Thus the
probability of reflection quickly increases in early times and we
find an exponential-law of expansion for our universe represented
by the 3d domain wall. This is an inflationary period of our
universe, as required by the current cosmology. The acceleration
saturates at very late time universe according to a quadratic
power-law of expansion. Accelerating universe conforms with
current observations. It is a very active subject in the modern
cosmology. Several scenarios addressing different problems in
accelerating universes such as brane/string gas cosmology and {\it
quartessence} have recently been considered in the literature ---
see e.g., \cite{brand} and \cite{ioav}, respectively. Our scenario
concerning 3d domain walls with internal structures colliding with
bulk particles can be considered as a complementary alternative to
braneworlds cosmology scenarios \cite{dgp,dg,defayet,ac-davis}.
This ``complementarity'' may shed some light on issues such as
accelerating universes in string/M-theory \cite{vijay,kklt,kklmmt}
and the cosmological constant problems \cite{weinberg,padna}. A
lot of issues concerning our scenario in brane cosmology of warped
5d space-time \cite{ac-davis} should also be investigated.

\acknowledgments

 We would like to thank D. Bazeia for discussions.
FFC and JFNO thank CNPq for fellowship.

\end{document}